%% file: main.tex
\documentclass[conference]{IEEEtran}
\IEEEoverridecommandlockouts
\usepackage{cite}
\usepackage{amsmath,amssymb,amsfonts}
\usepackage{algorithmic}
\usepackage{graphicx}
\usepackage{textcomp}
\usepackage{xcolor}
\usepackage{bm}
\usepackage{dsfont}
\usepackage{subfig}
\usepackage{hyperref}
\usepackage{tikz,tikz-3dplot}
\usetikzlibrary{calc, shapes.misc, intersections}
\usetikzlibrary{calc,shapes.misc}
\def\BibTeX{{\rm B\kern-.05em{\sc i\kern-.025em b}\kern-.08em
    T\kern-.1667em\lower.7ex\hbox{E}\kern-.125emX}}

\DeclareMathOperator{\cov}{cov}

\newtheorem{definition}{Definition}

\newcommand{\T}{\mathsf{T}}

\begin{document}

\title{A Proof of the Weak Simplex Conjecture\\
\thanks{This article is under peer review.}
}

\author{
\IEEEauthorblockN{Adriano Pastore}
\IEEEauthorblockA{
	Information and Signal Processing for Intelligent Communications \\
	Centre Tecnològic de Telecomunicacions de Catalunya (CTTC)\\
	Av.~Carl Friedrich Gauss 7, 08860 Castelldefels, Spain \\
	E-Mail: adriano.pastore@cttc.cat}
}

\maketitle

\begin{abstract}
We solve a long-standing open problem about the optimal codebook structure of codes in $n$-dimensional Euclidean space that consist of $n+1$ codewords subject to a per-codeword energy constraint, in terms of minimizing the average decoding error probability. The conjecture states that optimal codebooks consist of the $n+1$ vertices of a regular simplex (the $n$-dimensional generalization of a regular tetrahedron) inscribed in the unit sphere. A self-contained proof of this conjecture is provided that hinges on symmetry arguments and leverages a relaxation approach that consists in \emph{jointly} optimizing the codebook and the decision regions, rather than the codeword locations alone.
\end{abstract}


\section{Introduction}

The fundamental relationships between codebook geometry and decoding error probability in the classical Gaussian channel are at the core of some foundational information-theoric questions ever since the early days when Shannon constructed analytic and geometric approaches to address the problem of designing codes for communication in noise~\cite{Shannon1949} and to establish links between communication limits and sphere packing bounds under different types of power constraints~\cite{Shannon1959}. The case of $n+1$ codewords in $n$-dimensional Euclidean space is of special theoretic interest in that it corresponds to a configuration in which the pairwise codeword distance can be set constant (for any pair of codewords) when choosing the highly symmetric, regular simplex shape.

The origins of the simplex conjecture can be traced back to Shannon himself, who suspected that the maximization of minimum codeword distance, though sensible it may be as an approach to code design, might actually be suboptimal. Soon the conjecture emerged that, under a per-codeword energy constraint, the regular simplex inscribed in a centered $(n-1)$-sphere should indeed be optimal. Balakrishnan produced one of the first comprehensive studies of this conjecture~\cite{Balakrishnan1961}, but could only prove partial results. Dunbridge~\cite{Dunbridge1967} showed that the regular simplex is asymptotically optimal at low noise, and \emph{locally} optimal at every noise level.
Landau and Slepian reported a full proof~\cite{Landau1966} as early as 1966, based on geometric arguments involving the volume of the union of spherical caps, which later turned out to only hold true for the case $n=3$. The issue that prevented the Landau--Slepian approach from generalizing to arbitrary $n$ is discussed, e.g., in~\cite{Farber1968}.

The conjecture was further popularized by Tom Cover in 1987 in a famous compendium of open problems~\cite{Cover1987} as well as by James Massey's 1988 Shannon Lecture~\cite{Massey1988}. Massey suggested a distinction between two variants of the conjecture: the classic, so-called Weak Simplex Conjecture (WSC) in which codewords are subject to a bounded energy constraint, as well as a stronger claim, which relaxes this constraint to a bound on the sum-energy of all codewords. The latter variant of the conjecture gained the spotlight and became known as the Strong Simplex Conjecture (SSC). If proven true, the SSC would have settled the WSC too. However, the SSC was disproved by Steiner in 1994, who produced a counterexample in form of a signal set construction that is \emph{not} inscribed in the sphere and that outperforms the regular simplex~\cite{Steiner1994}.

This disproof settled the SSC in the negative, but left the WSC open.
Later analyses such as~\cite{Sun1997} accrued more evidence in favor of the WSC using numerical methods.
Balitskiy \emph{et al.} \cite{Balitskiy2017} gave a modern survey and provided some equivalent reformulations. Another recent paper, which discusses links to the closely related Simplex Mean Width conjecture, is~\cite{Litvak2018}.

Our proof starts with the original problem formulation and leverages a relaxation approach which, to the best of our knowledge, was not attempted by other authors.
Before delving into a formal proof of the general case, we will sketch the general intuition behind our proof method for the three-dimensional case ($n=3$).

\section{Proof sketch}

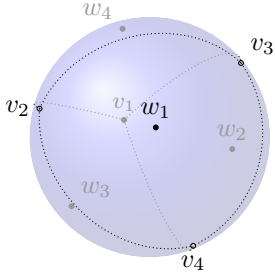
\begin{figure}[ht!]
	\centering
	\input{tikz/3d-sphere}
	\caption{A codebook $\mathcal{W} = (\bm{w}_1, \bm{w}_2, \bm{w}_3, \bm{w}_4)$ for $n=3$, located on the unit 2-sphere with decision regions delimited by points from $\mathcal{V} = (\bm{v}_1, \bm{v}_2, \bm{v}_3, \bm{v}_4)$ forming spherical triangles.}
	\label{fig:3D_codebook}
\end{figure}

Consider a problem of optimally placing four points $\mathcal{W} = (\bm{w}_1, \bm{w}_2, \bm{w}_3, \bm{w}_4)$ on the unit sphere $\mathbb{S}_2 \subset \mathbb{R}^3$.\footnote{For consistency we have chosen to denote collections of points as \emph{ordered} tuples throughout, although in many cases, the order does not matter. Instances where there is permutation invariance should be clear from context.} It is well known that the regular tetrahedron inscribed in the sphere is the configuration that maximizes the minimum distance between any two points. If $\bm{w}$ denotes one of the four points chosen uniformly at random and $\bm{z} \in \mathbb{R}^3$ stands for a normal Gaussian variable, then the WSC (for $n=3$) consists in the claim that the regular tetrahedron also maximizes the average probability of correctly guessing $\bm{w}$ from a noisy observation $\bm{y} = \bm{w} + \sigma\bm{z}$ (where $\sigma^2$ is the noise variance). That is, the regular tetrahedron maximizes
\begin{equation*}
Q(\mathcal{W})
	= \frac{1}{4} \sum_{i=1}^4 \mathsf{P}\Bigl\{ \lVert \bm{y} - \bm{w}_i \rVert \leq \min_j \lVert \bm{y} - \bm{w}_j \rVert \Big| \bm{w} = \bm{w}_i \Bigr\}.
\end{equation*}
This decoding metric naturally partitions $\mathbb{R}^3$ into four Voronoi regions $\mathcal{D}_1, \dotsc, \mathcal{D}_4$ that are flat-faceted cones with triangular section, whose apices are all located at the origin. These regions are delimited by bisecting planes that pass through the origin and whose intersections with $\mathbb{S}_2$ form a tiling of the sphere with four spherical triangles, the vertices of which we denote as $\mathcal{V} = (\bm{v}_1, \bm{v}_2, \bm{v}_3, \bm{v}_4)$. The codebook $\mathcal{W}$ thus fully determines the Voronoi cells $\mathcal{D}_i$ (equivalently, the triangle vertices $\mathcal{V}$) and vice-versa. Figure~\ref{fig:3D_codebook} provides an illustration.

Now consider a relaxation of the optimization problem where $\mathcal{V}$ and $\mathcal{W}$ are \emph{not} mutually determined: instead, we merely impose that every $\mathcal{D}_i$ (more specifically, its associated spherical triangle) contains a \emph{single} codeword $\bm{w}_i$. In such configurations, different codewords are \emph{not} necessarily mirror images of each other along triangle edges, because we no longer impose that the cones $\mathcal{D}_i$ be delimited by \emph{bisecting} planes, but instead by any set of planes passing through the origin that partition $\mathbb{R}^3$ into $4$ cones, each containing a single codeword.

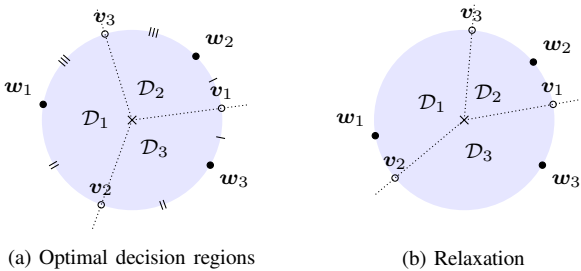
\begin{figure}[ht!]
\centering
\subfloat[][Optimal decision regions]{
	\input{tikz/2d-sphere_optimal_regions}
	\label{fig:optimal_decision_regions}
}
\hspace{0.2cm}
\subfloat[][Relaxation]{
	\input{tikz/2d-sphere_relaxation}
	\label{fig:relaxation}
}
\caption{Fig.~\ref{fig:optimal_decision_regions} shows a sketch (for the case $n=2$ for an easier visualization) of a codebook $\mathcal{W}$ with its optimal decision regions delimited by vectors in $\mathcal{V}^\ast(\mathcal{W})$. By contrast, Fig.~\ref{fig:relaxation} illustrates a valid pair $(\mathcal{W},\mathcal{V})$ under the relaxed assumptions.}
\end{figure}

That is, we relax the original optimization problem of the tetrahedron $\mathcal{W}$, i.e.,
\begin{equation*}
\max_{\mathcal{W}} Q(\mathcal{W})
	= \max_{\mathcal{W}} \frac{1}{4} \sum_{i=1}^4 \mathsf{P}\Bigl\{ \bm{y} \in \mathcal{D}_i \Big| \bm{w} = \bm{w}_i \Bigr\}
\end{equation*}
to one where we also co-optimize the tetrahedron $\mathcal{V}$, i.e.,
\begin{equation*}
\max_{\mathcal{W}, \mathcal{V}} Q(\mathcal{W},\mathcal{V})
	= \max_{\mathcal{W}, \mathcal{V}} \frac{1}{4} \sum_{i=1}^4 \mathsf{P}\Bigl\{ \bm{y} \in \mathcal{D}(\mathcal{V}_{\setminus i}) \Big| \bm{w} = \bm{w}_i \Bigr\}.
\end{equation*}
Here, $\mathcal{V}_{\setminus i}$ denotes the tuple $\mathcal{V}$ with the element $\bm{v}_i$ removed, and $\mathcal{D}(\mathcal{V}_{\setminus i})$ denotes the cone generated by $\mathcal{V}_{\setminus i}$.

The key reasoning behind proving that regular (equilateral) tetrahedra $\mathcal{W}, \mathcal{V}$ solve this double maximization is that, given a spherical triangle, e.g., formed by vertices $(\bm{v}_2, \bm{v}_3, \bm{v}_4)$, the placement of the associated codeword $\bm{w}_1$ on its interior is optimal at the point where $\mathcal{D}(\bm{v}_2, \bm{v}_3, \bm{v}_4)$ contains the maximum probability mass of a $\sigma^2$-variance Gaussian variable centered at $\bm{w}_1$. This optimal location, which we call the \emph{Gaussian centroid} of the spherical triangle formed by vertices $(\bm{v}_2, \bm{v}_3, \bm{v}_4)$, can be shown to be unique. In other words, the vertex tuple $(\bm{v}_2, \bm{v}_3, \bm{v}_4)$ uniquely determines (up to permutation) the optimal codeword $\bm{w}_1$. Likewise, using a proof by contradiction, we can show that fixing two vertices, e.g., $(\bm{v}_2, \bm{v}_3)$, as well as the centroid $\bm{w}_1$, uniquely determines the third vertex $\bm{v}_4$.

Now consider any two adjacent spherical triangles, e.g., $(\bm{v}_2, \bm{v}_3, \bm{v}_4)$ and $(\bm{v}_1, \bm{v}_2, \bm{v}_3)$. We know that in an \emph{optimal} configuration $(\mathcal{W}^\ast, \mathcal{V}^\ast)$, their respective codewords $\bm{w}_1^\ast$ and $\bm{w}_4^\ast$ must be mirrors of each other along the common edge (i.e., along the plane spanned by $(\bm{v}_2^\ast, \bm{v}_3^\ast)$ and the origin), and we also know that, as argued before, this common edge and the mirrored centroids $\bm{w}_1^\ast$ and $\bm{w}_4^\ast$ determine the vertices $\bm{v}_1^\ast$ and $\bm{v}_4^\ast$. This means that the spherical triangles $(\bm{v}_2^\ast, \bm{v}_3^\ast, \bm{v}_4^\ast)$ and $(\bm{v}_1^\ast, \bm{v}_2^\ast, \bm{v}_3^\ast)$ are reflections of each other along the common edge $(\bm{v}_2^\ast, \bm{v}_3^\ast)$. Since this reasoning applies to all pairs of triangles, we conclude that all four triangles that tile $\mathbb{S}_2$ are congruent and mutual reflections of each other, leading to $\mathcal{V}^\ast$ (and consequently $\mathcal{W}^\ast$) being a regular tetrahedron.

Yet another way of looking at the problem is that the relaxed double maximization could be addressed by an \emph{alternating} optimization of the centroids $\mathcal{W}$ (for fixed $\mathcal{V}$) and of the spherical triangles $\mathcal{V}$ (for fixed $\mathcal{W}$) in a way that is reminiscent of the Lloyd-Max algorithm. We prove by an argument of symmetry that regular tetrahedra are the only possible point of equilibrium.

\section{Formal proof}

Consider a spherical codebook $\mathcal{W} = ( \bm{w}_1, \dotsc, \bm{w}_{n+1} )$ of $n+1$ distinct unit-norm codewords from the $n$-dimensional Euclidean space $\mathbb{R}^n$ equipped with the standard scalar product $\langle \cdot , \cdot \rangle$. The codewords lie on the $(n-1)$-dimensional unit sphere\footnote{We will omit the subscript index used in the usual notation $\mathcal{S}_{n-1}$, since it will be $n-1$ in all cases.}
\begin{equation}
	\mathbb{S}
	= \bigl\{ \bm{w} \in \mathbb{R}^n \colon \lVert \bm{w} \rVert = 1 \bigr\}
\end{equation}
where $\lVert \cdot \rVert$ denotes the standard Euclidean norm. Let us further define the $n$-dimensional Gaussian measure $\gamma$ with density
\begin{equation}   \label{gaussian_density}
	g(\bm{x})
	= \frac{1}{(2\pi)^{n/2}} e^{-\frac{1}{2}\lVert \bm{x} \rVert^2}.
\end{equation}
Next we define a set of $n+1$ disjoint \emph{open} decision regions $\mathcal{D}_1, \dotsc, \mathcal{D}_{n+1}$ (where $\mathcal{D}_i$ contains codeword $\bm{w}_i$) whose closures $\overline{\mathcal{D}}_i = \mathcal{D}_i \cup \partial\mathcal{D}_i$ cover the Euclidean space $\mathbb{R}^n = \bigcup_{i=1}^{n+1} \overline{\mathcal{D}}_i$. Furthermore, we establish the decoding rule in such way that the decoder declares codeword $\bm{w}_i$ whenever the noisy vector $\bm{y} = \bm{w} + \sigma\bm{z}$ falls within the decision region $\mathcal{D}_i$. The average probability $Q(\mathcal{W})$ of correct decoding is then
\begin{equation}   \label{Q_W}
	Q(\mathcal{W})
	= \frac{1}{n+1} \sum_{i=1}^{n+1} \mathsf{P}\bigl\{ \bm{y} \in \mathcal{D}_i \big| \bm{w} = \bm{w}_i \bigr\}
\end{equation}
where each of the summands can be expressed as an integral
\begin{IEEEeqnarray}{rCl}
	\mathsf{P}\bigl\{ \bm{y} \in  \mathcal{D}_i \big| \bm{w} = \bm{w}_i \bigr\}
	&=& \frac{1}{\sigma^n} \int_{\mathcal{D}_i} g{\left(\frac{\bm{x} - \bm{w}_i}{\sigma}\right)} \mathrm{d}\bm{x}.
\end{IEEEeqnarray}

\subsection{Geometry of optimal decision regions}

As a consequence of the Gaussian density~\eqref{gaussian_density} being a monotone function of the norm $\lVert \bm{x} \rVert$, we have that, given any codebook $\mathcal{W}$, the optimal (open) decision regions $\mathcal{D}_i$, in terms of maximizing $Q(\mathcal{W})$, are the sets of points closer to $\bm{w}_i$ (in Euclidean distance) than to any other codeword. This well-known fact translates into $\mathcal{D}_i$ being open Voronoi cells pertaining respectively to codewords $\bm{w}_i$. Since the $\bm{w}_i$ lie on the unit sphere, each $\mathcal{D}_i$ is the intersection of $n$ open half spaces $\{ \bm{x} \in \mathbb{R}^n \colon \langle \bm{w}_i - \bm{w}_j , \bm{x} \rangle > 0 \}$ delimited by planes cutting through the origin. That is, the decision regions $\mathcal{D}_i$ can be represented as
\begin{IEEEeqnarray*}{rCl}
	\mathcal{D}_i
	&=& \left\{ \bm{x} \in \mathbb{R}^n \colon \bm{W}_i \bm{x} > \bm{0} \right\},
	\quad i \in [1 : n{+}1]   \IEEEeqnarraynumspace\IEEEyesnumber\label{D_i}
\end{IEEEeqnarray*}
where the rows of the square matrix $\bm{W}_i \in \mathbb{R}^{n \times n}$ are given by codeword differences $(\bm{w}_i - \bm{w}_j)^\T,\ j \in [1:n{+}1] \setminus \{i\}$ and the vector inequality in~\eqref{D_i} shall mean that inequality holds for all vector entries simultaneously. Here, $[1:n{+}1]$ denotes the integer range $[1 : n{+}1]$.

In the following, we will first assume that codewords are \emph{affinely independent}, i.e., that any matrix $\bm{W}_i, i = [1 : n{+}1]$ is full-rank.
Under this assumption, the parameter $\bm{x} \in \mathbb{R}^n$ in~\eqref{D_i} may be substituted by $\bm{\bar{\alpha}} \in \mathbb{R}^n$ via $\bm{x} = \bm{W}_i^{-1} \bm{\bar{\alpha}}$, which leads to a new parametrization of decision regions:
\begin{IEEEeqnarray*}{rCl}
	\mathcal{D}_i
	&=& \left\{ \bm{W}_i^{-1} \bm{\bar{\alpha}} \colon \bm{\bar{\alpha}} > \bm{0} \right\},
	\quad i \in [1 : n{+}1].
\end{IEEEeqnarray*}
If we normalize the columns of $\bm{W}_i^{-1}$ to unit length and denote the resulting column vectors as $\bm{v}_j^{(i)} \in \mathbb{S},\ j \in [1 : n{+}1] \setminus \{i\}$, then $\mathcal{D}_i$ can be simply represented as the cone generated by this collection of $n$ unit-length vectors. Furthermore, by rearranging equations it is easy to check that $\bm{W}_i^{-1}$ and $\bm{W}_{i'}^{-1}$ (with $i \neq i'$) have $n-1$ columns in common, which after normalization turn out to be pairwise equal, in such manner that $\bm{v}_j^{(i)} = \bm{v}_j^{(i')}$. Therefore, the superscript `$(i)$' can be dropped from notation, so we may define a tuple $\mathcal{V} = ( \bm{v}_1, \dotsc, \bm{v}_{n+1} )$ of unit vectors that constitute corner points delimiting the conical decision regions $\mathcal{D}_i = \mathcal{D}(\mathcal{V}_{\setminus i})$, namely,
\begin{IEEEeqnarray*}{rCl}
	\mathcal{D}(\mathcal{V}_{\setminus i})
	&=& \left\{ {\sum}_{j \neq i} \alpha_j \bm{v}_j \colon \bm{\alpha} > \bm{0} \right\},
	\quad i \in [1 : n{+}1]
\end{IEEEeqnarray*}
where $\bm{\alpha} \in \mathbb{R}_+^n$ is the vector of coordinates $\alpha_j,\ j \in [1 : n{+}1] \setminus \{i\}$ and $\mathcal{V}_{\setminus i}$ is shorthand for $\mathcal{V}$ with the $i$-th entry (i.e., $\bm{v}_i$) removed.

In conclusion, we can say that, given any tuple of $n+1$ affinely independent codewords $\mathcal{W}$, the \emph{optimal} tuple of $n$ so-called \emph{vertices} $\mathcal{V}$ delimiting the optimal decision regions $\mathcal{D}_i$, which we may as well denote as $\mathcal{V}^\ast(\mathcal{W})$, is fully and uniquely determined by $\mathcal{W}$ via the above described construction. Note that by this same construction, it is straightforward to conclude that the vectors in $\mathcal{V}^\ast(\mathcal{W})$ are affinely independent too.
By contrast, the case of \emph{affinely dependent} codebooks can be discarded without loss of generality, since affinely dependent codewords would entail linearly dependent $\mathcal{V}_{\setminus i}$, which corresponds to a suboptimal configuration according to Farber's argument laid out in~\cite[p.~11 and following]{Farber1968}.

\subsection{The relaxation approach}   \label{ssec:relaxation_approach}

Let us now consider a relaxation where $\mathcal{V}$ is \emph{not} determined by the codebook $\mathcal{W}$ in the way described in the previous section, i.e., we do not necessarily set $\mathcal{V}$ to be equal to $\mathcal{V}^\ast(\mathcal{W})$, hence the $\mathcal{D}_i$ (which is shorthand for $\mathcal{D}(\mathcal{V}_{\setminus i})$) are not necessarily chosen to be optimal decision regions. Instead, any pair $(\mathcal{W},\mathcal{V})$ may be chosen \emph{freely}, on the proviso that it retains the following properties, which are characteristic of any optimality candidate $(\mathcal{W}, \mathcal{V}^\ast(\mathcal{W}))$:
\begin{enumerate}
	\item	each spherical $(n-1)$-simplex formed by $n$ vertices $\mathcal{V}_{\setminus i}$ contains exactly one codeword $\bm{w}_i$ in its interior;
	\item	$\mathcal{W}$ and $\mathcal{V}$ are sets of affinely independent vectors.
\end{enumerate}
With any valid pair $(\mathcal{W}, \mathcal{V})$ satisfying these two constraints, we can rewrite the maximization
objective as
\begin{equation}   \label{Q_WV}
	Q(\mathcal{W}, \mathcal{V})
	= \frac{1}{n+1} \sum_{i=1}^{n+1} \mathsf{P}\bigl\{ \bm{y} \in \mathcal{D}(\mathcal{V}_{\setminus i}) \big| \bm{w} = \bm{w}_i \bigr\}.
\end{equation}
The relaxation of the original problem $\max_{\mathcal{W}} Q(\mathcal{W})$ now consists in solving $\max_{(\mathcal{W}, \mathcal{V})} Q(\mathcal{W}, \mathcal{V})$ over all pairs of compatible $\mathcal{W}$ and $\mathcal{V}$.
So far, we have elicited that any jointly optimal pair $(\mathcal{W}^\ast, \mathcal{V}^\ast)$ of the relaxed problem must satisfy the marginal optimality $\mathcal{V}^\ast = \mathcal{V}^\ast(\mathcal{W}^\ast)$ where the optimal vertex construction $\mathcal{V}^\ast(\mathcal{W})$ is as discussed earlier. Next, to further characterize some key properties of any maximizing pair $(\mathcal{W}^\ast, \mathcal{V}^\ast)$, we move on to discuss properties of the marginally optimal codebook construction $\mathcal{W}^\ast(\mathcal{V})$. For this purpose, we introduce the concept of \emph{Gaussian centroid}.
\begin{definition}   \label{def:gaussian_centroid}
A (spherical) Gaussian centroid $\bm{c} \in \mathbb{S}$ of a measurable spherical set $\mathcal{S} \subset \mathbb{S}$ with parameter $\sigma > 0$ is a unit-norm vector at which a Gaussian variable $\bm{g}_{\bm{c}} \sim \mathcal{N}(\bm{c}, \sigma^2\mathbf{I}_n)$ is centered so as to maximize the probability $\mathsf{P}\{ \bm{g}_{\bm{c}} \in \mathcal{C}(\mathcal{S}) \}$, where $\mathcal{C}(\mathcal{S}) = \{ \alpha \bm{s} \colon \alpha \geq 0, \bm{s} \in \mathcal{S} \}$ denotes the cone generated by $\mathcal{S}$.
\end{definition}
We are now ready to articulate two \emph{necessary} conditions that any jointly optimal pair $(\mathcal{W}^\ast,\mathcal{V}^\ast)$ must necessarily satisfy:
\begin{enumerate}
    \item   For any fixed $\mathcal{W}$, the marginally optimal $\mathcal{V}^\ast(\mathcal{W})$ is such that any hyperplane spanned by a sub-tuple $\mathcal{V}^\ast(\mathcal{W})_{\setminus i, \setminus j}$ bisects the connecting segment between $\bm{w}_i$ and $\bm{w}_j$. In other words, $\bm{w}_i$ and $\bm{w}_j$ are mirror images of each other with respect to said bisecting hyperplane.
    \item   For any fixed $\mathcal{V}$, the marginally optimal $\mathcal{W}^\ast(\mathcal{V})$ is such that for each cone $\mathcal{D}(\mathcal{V}_{\setminus i})$, the corresponding codeword $\bm{w}_i \in \mathcal{D}(\mathcal{V}_{\setminus i})$ is a \emph{Gaussian centroid} of the spherical $(n-1)$-simplex $\mathcal{D}(\mathcal{V}_{\setminus i}) \cap \mathbb{S}$.
\end{enumerate}
The first of these conditions is well known and was discussed in detail in the previous section as an immediate consequence of the optimality of nearest-neighbor decoding. The second follows straightforwardly from Definition~\ref{def:gaussian_centroid}, since the placement of each codeword $\bm{w}_i$ at a centroid location guarantees that the $i$-th probability term in the sum \eqref{Q_WV} is maximized, without impact on the other summands.

\subsection{Uniqueness of Gaussian centroids}   \label{sec:uniqueness}

We will show that any spherical $n$-simplex has a \emph{unique} Gaussian centroid, which is thus fully determined by the vertices of that simplex.
First, recall that a Gaussian centroid of $\mathcal{D}(\mathcal{V}_{\setminus i}) \cap \mathbb{S}$ does by definition maximize the integral expression
\begin{IEEEeqnarray}{rCl}
	Q_i(\bm{c})
	&=& \frac{1}{\sigma^n} \int_{\mathcal{D}_i} g{\left( \frac{\bm{x} - \bm{c}}{\sigma} \right)} \mathrm{d}\bm{x}
\end{IEEEeqnarray}
subject to $\lVert \bm{c} \rVert = 1$.
Hence the gradient
\begin{IEEEeqnarray*}{rCl}
	\nabla Q_i(\bm{c})
	&=& \frac{1}{\sigma^n} \int_{\mathcal{D}_i} \nabla_{\!\bm{c}} \, g{\left( \frac{\bm{x} - \bm{c}}{\sigma} \right)} \mathrm{d}\bm{x} \\
	&=& \frac{1}{\sigma^{n+2}} \int_{\mathcal{D}_i} (\bm{x} - \bm{c}) g{\left( \frac{\bm{x} - \bm{c}}{\sigma} \right)} \mathrm{d}\bm{x} \\
	&=& \frac{\mathsf{P}\{ \bm{g}_{\bm{c}} \in \mathcal{D}_i \}}{\sigma^2} \Bigl( \mathsf{E}\bigl[ \bm{g}_{\bm{c}} \big| \bm{g}_{\bm{c}} \in \mathcal{D}_i \bigr] - \bm{c} \Bigr)
\end{IEEEeqnarray*}
where $\bm{g}_{\bm{c}}$ stands for a $\bm{c}$-centered Gaussian $\bm{g}_{\bm{c}} \sim \mathcal{N}(\bm{c},\sigma^2 \mathbf{I}_n)$, must be collinear with $\bm{c}$ at a centroid location (i.e., the gradient has only radial, but no tangential components on the sphere at a centroid location), which in turn implies that $\mathsf{E}\bigl[ \bm{g}_{\bm{c}} \big| \bm{g}_{\bm{c}} \in \mathcal{D}_i \bigr]$ must be collinear with $\bm{c}$. Factoring in the unit-norm constraint $\lVert \bm{c} \rVert = 1$, we conclude that an optimal $\bm{c}$ must satisfy the fixed-point equation $\bm{c} = \mathbf{f}_i(\bm{c})$, where
\begin{equation}   \label{fixed-point_function}
	\mathbf{f}_i \colon \
	\mathbb{S} \to \mathbb{S}, \
	\bm{c} \mapsto \frac{\mathsf{E}\bigl[ \bm{g}_{\bm{c}} \big| \bm{g}_{\bm{c}} \in \mathcal{D}_i \bigr]}{\big\lVert \mathsf{E}\bigl[ \bm{g}_{\bm{c}} \big| \bm{g}_{\bm{c}} \in \mathcal{D}_i \bigr] \big\rVert}.
\end{equation}
To prove that any fixed-point iteration $\bm{c}_{m+1} = \mathbf{f}_i(\bm{c}_m), m \in \mathbb{N}$ converges, we proceed to showing that the self-map $\mathbf{f}_i$ is contractive. For this we represent $\mathbf{f}_i = \mathbf{g} \circ \mathbf{h}_i$ as the concatenation of $\mathbf{g}(\bm{x}) = \bm{x} / \lVert \bm{x} \rVert$ and $\mathbf{h}_i(\bm{c}) = \mathsf{E}\bigl[ \bm{g}_{\bm{c}} \big| \bm{g}_{\bm{c}} \in \mathcal{D}_i \bigr]$, such that by the chain rule, the Jacobian
\begin{equation}   \label{Jacobian_product_rule}
	\bm{J}\mathbf{f}_i(\bm{c})
	= \bm{J}(\mathbf{g} \circ \mathbf{h}_i)(\bm{c})
	= \bm{J}\mathbf{g}( \mathbf{h}_i(\bm{c}) ) \cdot \bm{J}\mathbf{h}_i(\bm{c})
\end{equation}
factorizes as the matrix product of two Jacobians, namely
\begin{subequations}
\begin{IEEEeqnarray}{rCl}
	\bm{J}\mathbf{g}(\bm{x})
	&=& \frac{1}{\lVert \bm{x} \rVert} \left( \mathbf{I} - \frac{\bm{x}}{\lVert \bm{x} \rVert} \frac{\bm{x}^\T}{\lVert \bm{x} \rVert} \right)   \label{J1} \\
	\bm{J}\mathbf{h}_i(\bm{c})
	&=& -\frac{1}{\sigma^2} \cov(\bm{g}_{\bm{c}} | \bm{g}_{\bm{c}} \in \mathcal{D}_i).   \label{J2}
\end{IEEEeqnarray}
\end{subequations}
Expression~\eqref{J2} can be derived after some algebra. Indeed, if we denote the measure of $\bm{g}_{\bm{c}} \sim \mathcal{N}(\bm{c},\sigma^2 \mathbf{I}_n)$ as $\gamma_{\bm{c}}$ and abbreviate its differential element as $\mathrm{d}\gamma_{\bm{c}}(\bm{x}) = g{\left( \frac{\bm{x} - \bm{c}}{\sigma} \right)} \mathrm{d}\bm{x}$,
\begin{IEEEeqnarray*}{rCl}
	\IEEEeqnarraymulticol{3}{l}{
		\bm{J}_{\bm{c}} \mathsf{E}\bigl[ \bm{g}_{\bm{c}} \big| \bm{g}_{\bm{c}} \in \mathcal{D}_i \bigr]
	} \\ \ 
	&=& \frac{1}{\sigma^n \gamma_{\bm{c}}(\mathcal{D}_i)} \int_{\mathcal{D}_i} \bm{x} \bm{\nabla}_{\!\bm{c}}^\T \mathrm{d}\gamma_{\bm{c}}(\bm{x}) \\
	&& {} - \frac{1}{\sigma^{2n} \gamma_{\bm{c}}(\mathcal{D}_i)^2} \int_{\mathcal{D}_i} \bm{x} \mathrm{d}\gamma_{\bm{c}}(\bm{x}) \int_{\mathcal{D}_i} \bm{\nabla}_{\!\bm{c}}^\T \mathrm{d}\gamma_{\bm{c}}(\bm{x}) \\
	&=& -\frac{1}{\sigma^{n+2} \gamma_{\bm{c}}(\mathcal{D}_i)} \int_{\mathcal{D}_i} \bm{x} (\bm{x} - \bm{c})^\T \mathrm{d}\gamma_{\bm{c}}(\bm{x}) \\
	&& {} + \frac{1}{\sigma^{2n+2} \gamma_{\bm{c}}(\mathcal{D}_i)^2} \int_{\mathcal{D}_i} \bm{x} \mathrm{d}\gamma_{\bm{c}}(\bm{x}) \int_{\mathcal{D}_i} (\bm{x} - \bm{c})^\T \mathrm{d}\gamma_{\bm{c}}(\bm{x}) \\
	&=& -\frac{1}{\sigma^2} \Bigl( \mathsf{E}\bigl[ \bm{g}_{\bm{c}} \bm{g}_{\bm{c}}^\T \big| \bm{g}_{\bm{c}} \in \mathcal{D}_i \bigr] - \mathsf{E}\bigl[ \bm{g}_{\bm{c}} \big| \bm{g}_{\bm{c}} \in \mathcal{D}_i \bigr] \bm{c}^\T \Bigr) \\
	&& {} + \frac{1}{\sigma^2} \Bigl( \mathsf{E}\bigl[ \bm{g}_{\bm{c}} \big| \bm{g}_{\bm{c}} \in \mathcal{D}_i \bigr] \bigl( \mathsf{E}\bigl[ \bm{g}_{\bm{c}}^\T \big| \bm{g}_{\bm{c}} \in \mathcal{D}_i \bigr] - \bm{c}^\T \bigr) \Bigr) \\
	&=& -\frac{1}{\sigma^2} \Bigl( \mathsf{E}\bigl[ \bm{g}_{\bm{c}} \bm{g}_{\bm{c}}^\T \big| \bm{g}_{\bm{c}} \in \mathcal{D}_i \bigr] + \mathsf{E}\bigl[ \bm{g}_{\bm{c}} \big| \bm{g}_{\bm{c}} \in \mathcal{D}_i \bigr] \mathsf{E}\bigl[ \bm{g}_{\bm{c}}^\T \big| \bm{g}_{\bm{c}} \in \mathcal{D}_i \bigr] \Bigr)
\end{IEEEeqnarray*}
whence~\eqref{J2}.

From~\cite[Cor.~2.1]{Kanter1977} and the convexity of $\mathcal{D}_i$ we infer that $\cov(\bm{g}_{\bm{c}}) - \cov(\bm{g}_{\bm{c}} | \bm{g}_{\bm{c}} \in \mathcal{D}_i)$ is positive semidefinite. Since $\cov(\bm{g}_{\bm{c}}) = \sigma^2 \mathbf{I}$, we further conclude that $\bm{J}\mathbf{h}_i(\bm{c})$ has eigenvalues with absolute value no larger than unity. Likewise, the projector $\mathbf{I} - \bm{x}\bm{x}^\T/\lVert \bm{x} \rVert^2$ appearing in~\eqref{J1} has eigenvalues confined on the unit interval.
To finish proving that the product of Jacobians $\bm{J}\mathbf{f}_i(\bm{c})$ [cf.~\eqref{Jacobian_product_rule}] is a (strictly) contractive mapping, we will show that the scalar factor in~\eqref{J1} is smaller than one. That is, we will show that $\big\lVert \mathsf{E}\bigl[ \bm{g}_{\bm{c}} \big| \bm{g}_{\bm{c}} \in \mathcal{D}_i \bigr] \big\rVert > 1$.

It will suffice to prove the stronger result $\mathsf{E}\bigl[ \bm{c}^\T \bm{g}_{\bm{c}} \big| \bm{g}_{\bm{c}} \in \mathcal{D}_i \bigr] > 1$ (recall that $\lVert \bm{c} \rVert = 1$). Note that $\mathcal{V}_{\setminus i}$ is a collection of $n$ linearly independent unit vectors, since the $n+1$ vectors in $\mathcal{V}$ are assumed affinely independent.
Notice also that it suffices to consider $\bm{c}$ belonging to the \emph{interior} of the cone $\mathcal{D}_i = \mathcal{D}(\mathcal{V}_{\setminus i})$, i.e., $\bm{c} = \sum_{j=1}^n \beta_i \bm{v}_j$ with positive coefficients $\beta_i > 0$.
As a consequence, any subset of size $n-1$ of the projected vectors $\bm{\Pi}_{\bm{c}} \bm{v}_j, j=1,\dotsc,n$ (where $\bm{\Pi}_{\bm{c}} = \mathbf{I} - \bm{c}\bm{c}^\T$) are linearly independent among them. Therefore, the matrix $\bm{A}^{-1} = \begin{bmatrix} \bm{c} & \bm{\Pi}_{\bm{c}} \bm{v}_2 & \hdots & \bm{\Pi}_{\bm{c}} \bm{v}_n \end{bmatrix}^\T \in \mathbb{R}^{n \times n}$ (or in short $\bm{A}^{-1} = \begin{bmatrix} \bm{c} & \bm{\Pi}_{\bm{c}} \bm{V}_1 \end{bmatrix}^\T$ with a full column-rank $\bm{\Pi}_{\bm{c}}\bm{V}_1 \in \mathbb{R}^{n \times (n-1)}$) is invertible. Let us define the bijective linear transformations
\begin{align}   \label{linear_transformation}
	\bm{u}(\bm{x})
	&= \bm{A}^{-1} \frac{\bm{x} - \bm{c}}{\sigma}
	&
	\bm{x}(\bm{u})
	&= \sigma \bm{A}\bm{u} + \bm{c}
\end{align}
such that, by a change of integration variable,
\begin{IEEEeqnarray*}{rCl}
		\mathsf{E}\bigl[ \bm{g}_{\bm{c}} \big| \bm{g}_{\bm{c}} \in \mathcal{D}_i \bigr]
	&=& \frac{ \int_{\mathbb{R}^n} \bm{x}(\bm{u}) g{(\bm{A}\bm{u})} \mathds{1}\bigl\{ \bm{x}(\bm{u}) \in \mathcal{D}_i \bigr\} \mathrm{d}\bm{u} }{ \int_{\mathbb{R}^n} g{(\bm{A}\bm{u})} \mathds{1}\bigl\{ \bm{x}(\bm{u}) \in \mathcal{D}_i \bigr\} \mathrm{d}\bm{u} }
\end{IEEEeqnarray*}
where $\mathds{1}\{\cdot\}$ denotes the indicator function.
Using the latter expression and~\eqref{linear_transformation}, we see that proving $\mathsf{E}\bigl[ \bm{c}^\T\bm{g}_{\bm{c}} \big| \bm{g}_{\bm{c}} \in \mathcal{D}_i \bigr] > 1$ boils down to proving
\begin{equation}   \label{quod_est}
	\int_{\mathbb{R}^n} \bm{c}^\T\bm{A}\bm{u} \, g{(\bm{A}\bm{u})} \, \mathds{1}\bigl\{ \bm{x}(\bm{u}) \in \mathcal{D}_i \bigr\} \mathrm{d}\bm{u}
	> 0.
\end{equation}
Note that $\bm{c}^\T\bm{A}\bm{u} = u_1$. This is because, given how $\bm{A}^{-1}$ was defined, we have $\bm{A} = \begin{bmatrix} \bm{c} & \bm{\Pi}_{\bm{c}} \bm{V}_1 ( \bm{V}_1^\T \bm{\Pi}_{\bm{c}} \bm{V}_1 )^{-1} \end{bmatrix}$, due to which $\bm{c}^\T\bm{A}$ is the first canonical basis (row) vector. Furthermore, if we denote by $\bm{u}^-$ the vector $\bm{u}$ with only the sign of its first entry $u_1$ flipped, then note that the Gaussian density term satisfies $g{(\bm{A}\bm{u})} = g{(\bm{A}\bm{u}^-)}$, because $g{(\bm{A}\bm{u})} = (2\pi)^{-n} \exp\bigl(-\bm{u}^\T\bm{A}^\T\bm{A}\bm{u}/2\bigr)$ depends on $u_1$ only via its square $u_1^2$, due to
\begin{equation}
	\bm{A}^\T\bm{A}
	= \begin{bmatrix} 1 & \bm{0}_{1 \times (n-1)} \\ \bm{0}_{(n-1) \times 1} & (\bm{V}_1^\T \bm{\Pi}_{\bm{c}} \bm{V}_1)^{-1} \end{bmatrix}.
\end{equation}
In~\eqref{quod_est}, upon splitting the integration over $u_1$ between positive and negative half axes $(-\infty,0]$ and $[0,+\infty)$, then flipping the integration domain of the first integral, $(-\infty,0]$, by substitution into $[0,+\infty)$, then eventually reassembling both integrals into one, we can turn the left-hand side of~\eqref{quod_est} into
\begin{multline}
	\int_{\mathbb{R}^n} u_1 g{(\bm{A}\bm{u})} \mathds{1}\bigl\{ \bm{x}(\bm{u}) \in \mathcal{D}_i \bigr\} \mathrm{d}\bm{u} \\
	= \int_0^\infty \int_{-\infty}^\infty \dotsi \int_{-\infty}^\infty u_1 g{(\bm{A}\bm{u})} \nu(\bm{u}) \, \mathrm{d}u_n \mathrm{d}u_{n-1} \dotso \mathrm{d}u_1   \label{nested_integral}
\end{multline}
where $\nu(\bm{u}) = \mathds{1}\bigl\{ \bm{x}(\bm{u}) \in \mathcal{D}_i \bigr\} - \mathds{1}\bigl\{ \bm{x}(\bm{u}^-) \in \mathcal{D}_i \bigr\}$. The function $\nu(\bm{u})$ is non-negative because $\bm{x}(\bm{u}) - \bm{x}(\bm{u}^-) = \sigma\bm{A}(\bm{u} - \bm{u}^-) = 2\sigma\bm{c} u_1$ is a non-negative multiple of $\bm{c}$, and therefore $\bm{x}(\bm{u}^-) \in \mathcal{D}_i$ implies $\bm{x}(\bm{u}) \in \mathcal{D}_i$, due to the conical shape of $\mathcal{D}_i$ and the fact that $\bm{c} \in \mathcal{D}_i$. It immediately follows that~\eqref{quod_est} holds, yet with a \emph{weak} inequality.

To strengthen this to a \emph{strict} inequality as stated in~\eqref{quod_est}, let us run a closer analysis of the term $\nu(\bm{u})$. Consider the shifted cone $\hat{\mathcal{D}}_i = \mathcal{D}_i - 2\sigma\bm{c} u_1$. Since for $u_1 > 0$ the shift vector $2\sigma\bm{c} u_1$ lies in the \emph{interior} of the original cone $\mathcal{D}_i$, it is straightforward to deduce that $\mathcal{D}_i \subset \hat{\mathcal{D}}_i$, where the set difference $\hat{\mathcal{D}}_i \setminus \mathcal{D}_i$ has positive Gaussian measure.

\begin{figure}[ht!]
	\centering
	\input{tikz/cone}
	\caption{Cone $\mathcal{D}_i$ with its shifted version $\hat{\mathcal{D}}_i$}
	\label{fig:shifted_cone}
\end{figure}
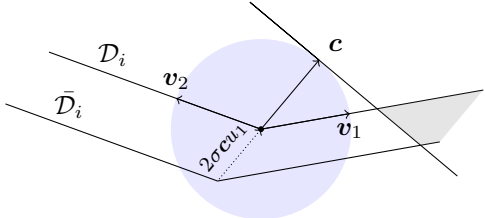

Reverting the previous substitution by means of~\eqref{linear_transformation} on the right-hand side of~\eqref{nested_integral} yields
\begin{multline}   \label{shifted_cone_integral}
	\int_{\mathbb{R}^n} u_1 g{(\bm{A}\bm{u})} \mathds{1}\bigl\{ \bm{x}(\bm{u}) \in \mathcal{D}_i \bigr\} \mathrm{d}\bm{u}
	= \frac{1}{\sigma |\det(\bm{A})|} \times {} \\
	{} \times \int_{\mathbb{R}^n} (\bm{c}^\T\bm{x} - 1) \mathds{1}\{ \bm{c}^\T\bm{x} \geq 1 \} g{\left( \frac{\bm{x} - \bm{c}}{\sigma} \right)} \nu(\bm{u}(\bm{x})) \, \mathrm{d}\bm{x}.
\end{multline}
Under the assumption that the cone $\mathcal{D}_i$ is strictly inscribed in a half space (which we may assume thanks to Balakrishnan's argument in~\cite[Sec.~II]{Balakrishnan1961}), the subset of $\mathbb{R}^n$ on which the product $\mathds{1}\{ \bm{c}^\T\bm{x} \geq 1 \} \nu(\bm{u}(\bm{x}))$ is equal to one, namely the intersection $( \hat{\mathcal{D}}_i \setminus \mathcal{D}_i ) \cap \{ \bm{x} \in \mathbb{R}^n \colon \bm{c}^\T\bm{x} \geq 1 \}$, as illustrated by the shaded area in Fig.~\ref{fig:shifted_cone}, has non-zero volume. Hence~\eqref{shifted_cone_integral} is positive.

This proves~\eqref{quod_est} and consequently, that $\mathbf{f}_i$ is a (strictly) contractive mapping on the compact closure of $\mathcal{D}_i \cap \mathbb{S}$. By Banach's fixed-point theorem, the equation $\bm{c} = \mathbf{f}_i(\bm{c})$ has a unique solution, which we further know to be an interior point of $\mathcal{D}_i \cap \mathbb{S}$. Therefore, the so-called \emph{Gaussian centroid} of $\mathcal{D}_i \cap \mathbb{S}$ is unique and the optimal codeword $\bm{w}_i$ for a given $\mathcal{V}_{\setminus i}$ is precisely that Gaussian centroid.

\subsection{One-to-one relationship between vertices and centroids}   \label{ssec:injectivity}

As was established in the previous section, the Gaussian centroid of spherical $n$-simplices is unique, so we are legitimized to define a Gaussian centroid function
\begin{equation}
	\bm{\zeta} \colon \underbrace{ \mathbb{S} \times \dotso \times \mathbb{S} }_{\text{$n$ factors}} \to \mathbb{S}
\end{equation}
which maps an $n$-tuple of linearly independent unit-norm vertices $\tilde{\mathcal{V}} = (\tilde{\bm{v}}_1, \dotsc, \tilde{\bm{v}}_n)$ to the Gaussian centroid $\bm{\zeta}(\tilde{\mathcal{V}})$ of the spherical $n$-simplex spanned by these vertices.
We now ask the following question: when $n-1$ vertices out of the $n$-tuple $\tilde{\mathcal{V}}$ are fixed, as well as the location of the Gaussian centroid $\bm{\zeta}$, is the remaining vertex fully determined?
To answer this question, consider the \emph{marginal} centroid function obtained from $\bm{\zeta}(\cdot)$ when fixing all but one argument. Since $\bm{\zeta}$ is symmetric, without loss of generality we may consider the last $n-1$ arguments $\tilde{\mathcal{V}}_{\setminus 1} = (\tilde{\bm{v}}_2, \dotsc, \tilde{\bm{v}}_n)$ to be fixed as parameters and define the marginal centroid function via
\begin{equation}
	\bm{\zeta}(\cdot ; \tilde{\mathcal{V}}_{\setminus 1}) \colon \ \mathbb{S} \to \mathbb{S}, \ \tilde{\bm{v}}_1 \mapsto \bm{\zeta}(\tilde{\mathcal{V}}).
\end{equation}
We are essentially asking whether this self-map is injective for every linearly independent $\tilde{\mathcal{V}}_{\setminus 1}$.
First, let $\tilde{\bm{e}}$ denote a unit vector generating the null space of $\tilde{\mathcal{V}}_{\setminus 1}$. Without loss of generality, we can restrict the domain and co-domain of $\bm{\zeta}(\cdot ; \tilde{\mathcal{V}}_{\setminus 1})$ to the open half-sphere $\mathbb{S} \cap \mathcal{H}(\tilde{\bm{e}})$ where $\mathcal{H}(\tilde{\bm{e}}) = \{ \bm{x} \colon \langle \tilde{\bm{e}}, \bm{x} \rangle > 0 \}$, because it is straightforward to see that the image and pre-image of $\bm{\zeta}(\cdot ; \tilde{\mathcal{V}}_{\setminus 1})$ will either both lie in $\mathcal{H}(\tilde{\bm{e}})$ or in $\mathcal{H}(-\tilde{\bm{e}})$. All subsequent reasoning holds for $\tilde{\bm{e}}$ and $-\tilde{\bm{e}}$ alike.

Hence, given two distinct pre-images $\tilde{\bm{v}}_1$ and $\tilde{\bm{v}}_1'$ ($\tilde{\bm{v}}_1 \neq \tilde{\bm{v}}_1'$) of the \emph{restricted} self-map $\bar{\bm{\zeta}}(\cdot ; \tilde{\mathcal{V}}_{\setminus 1}) \colon \mathbb{S} \cap \mathcal{H}(\tilde{\bm{e}}) \to \mathbb{S} \cap \mathcal{H}(\tilde{\bm{e}})$, we will argue that their images---which for notational brevity we denote as $\bm{c} = \bm{\zeta}(\tilde{\bm{v}}_1 ; \tilde{\mathcal{V}}_{\setminus 1})$ and $\bm{c}' = \bm{\zeta}(\tilde{\bm{v}}_1' ; \tilde{\mathcal{V}}_{\setminus 1})$, respectively---are necessarily different.
Let us also denote the cones $\tilde{\mathcal{D}} = \mathcal{D}(\tilde{\mathcal{V}})$ and $\tilde{\mathcal{D}}' = \mathcal{D}(\tilde{\mathcal{V}}')$ as well as their intersection $\mathcal{I} = \tilde{\mathcal{D}} \cap \tilde{\mathcal{D}}'$. We further introduce
\begin{subequations}
\begin{IEEEeqnarray}{rCll}
	\bm{i}
	&=& \mathsf{E}[\bm{g}_{\bm{c}} | \bm{g}_{\bm{c}} \in \mathcal{I}] &   \label{def:i} \\
	\bm{i}'
	&=& \mathsf{E}[\bm{g}_{\bm{c}'} | \bm{g}_{\bm{c}'} \in \mathcal{I}] &   \label{def:i_prime} \\
	\bm{t}
	&=& \mathsf{E}[\bm{g}_{\bm{c}} | \bm{g}_{\bm{c}} \in \tilde{\mathcal{D}} \setminus \tilde{\mathcal{D}}'] & \ \text{if $\tilde{\mathcal{D}} \setminus \tilde{\mathcal{D}}' \neq \emptyset$, else $\bm{t} = \bm{0}$}    \label{def:t} \\
	\bm{t}'
	&=& \mathsf{E}[\bm{g}_{\bm{c}'} | \bm{g}_{\bm{c}'} \in \tilde{\mathcal{D}}' \setminus \tilde{\mathcal{D}}] & \ \text{if $\tilde{\mathcal{D}}' \setminus \tilde{\mathcal{D}} \neq \emptyset$, else $\bm{t}' = \bm{0}$}.   \IEEEeqnarraynumspace\label{def:t_prime}
\end{IEEEeqnarray}
\end{subequations}
As previously, we denote the measure of $\bm{g}_{\bm{c}} \sim \mathcal{N}(\bm{c},\sigma^2 \mathbf{I}_n)$ as $\gamma_{\bm{c}}$.
Based on the fixed-point identity that any centroid must satisfy [cf.~Section~\ref{sec:uniqueness}], $\bm{c}$ and $\bm{c}'$ can be expressed as linear combinations of vectors $\bm{i}$, $\bm{t}$ and $\bm{t}'$:
\begin{subequations}
\begin{IEEEeqnarray}{rCCCl}
	\bm{c}
	&=& \frac{ \mathsf{E}[\bm{g}_{\bm{c}} | \bm{g}_{\bm{c}} \in \tilde{\mathcal{D}} ] }{ \lVert \mathsf{E}[\bm{g}_{\bm{c}} | \bm{g}_{\bm{c}} \in \tilde{\mathcal{D}} ] \rVert }
	&=& \frac{ \gamma_{\bm{c}}(\mathcal{I}) \bm{i} + \gamma_{\bm{c}}(\tilde{\mathcal{D}} \setminus \tilde{\mathcal{D}}') \bm{t} }{\lVert \gamma_{\bm{c}}(\mathcal{I}) \bm{i} + \gamma_{\bm{c}}(\tilde{\mathcal{D}} \setminus \tilde{\mathcal{D}}') \bm{t} \rVert}   \label{c} \\
	\bm{c}'
	&=& \frac{ \mathsf{E}[\bm{g}_{\bm{c}'} | \bm{g}_{\bm{c}'} \in \tilde{\mathcal{D}}' ] }{ \lVert \mathsf{E}[\bm{g}_{\bm{c}'} | \bm{g}_{\bm{c}'} \in \tilde{\mathcal{D}}' ] \rVert }
	&=& \frac{ \gamma_{\bm{c}'}(\mathcal{I}) \bm{i}' + \gamma_{\bm{c}'}(\tilde{\mathcal{D}}' \setminus \tilde{\mathcal{D}}) \bm{t}' }{\lVert \gamma_{\bm{c}'}(\mathcal{I}) \bm{i}' + \gamma_{\bm{c}'}(\tilde{\mathcal{D}}' \setminus \tilde{\mathcal{D}}) \bm{t}' \rVert}.   \IEEEeqnarraynumspace\label{c_prime}
\end{IEEEeqnarray}
\end{subequations}
We now perform a proof by contradiction: let us assume that $\tilde{\bm{v}}_1 \neq \tilde{\bm{v}}_1'$ and $\bm{c} = \bm{c}'$. From the former assumption ($\tilde{\bm{v}}_1 \neq \tilde{\bm{v}}_1'$), it follows that $\tilde{\mathcal{D}} \neq \tilde{\mathcal{D}}'$ and that at least one of the two cone differences $\tilde{\mathcal{D}} \setminus \tilde{\mathcal{D}}'$ or $\tilde{\mathcal{D}}' \setminus \tilde{\mathcal{D}}$ in non-empty hence at least one of either $\gamma_{\bm{c}}(\tilde{\mathcal{D}} \setminus \tilde{\mathcal{D}}')$ or $\gamma_{\bm{c}'}(\tilde{\mathcal{D}}' \setminus \tilde{\mathcal{D}})$ is positive. Meanwhile, the assumption $\bm{c} = \bm{c}'$ also implies $\bm{i} = \bm{i}'$ [cf.~\eqref{def:i}--\eqref{def:i_prime}] as well as $\gamma_{\bm{c}}(\mathcal{I}) = \gamma_{\bm{c}'}(\mathcal{I})$. As a result, we can replace $\bm{c}'$ by $\bm{c}$ and $\bm{i}'$ by $\bm{i}$ in \eqref{c}--\eqref{c_prime}.
We distinguish two cases:
\begin{itemize}
	\item	If $\bm{i}$, $\bm{t}$ and $\bm{t'}$ are linearly \emph{independent}, then the right-hand sides of~\eqref{c}--\eqref{c_prime} clearly cannot coincide, since we just saw that either $\gamma_{\bm{c}}(\tilde{\mathcal{D}} \setminus \tilde{\mathcal{D}}') \neq 0$ or $\gamma_{\bm{c}'}(\tilde{\mathcal{D}}' \setminus \tilde{\mathcal{D}}) \neq 0$.
	\item	If $\bm{i}$, $\bm{t}$ and $\bm{t'}$ are linearly \emph{dependent}, we claim that this can only occur if $\bm{t} = \bm{0}$ (resp. $\bm{t}' = \bm{0}$) and $\bm{i}, \bm{t}'$ (resp. $\bm{i}, \bm{t}$) are non-collinear, or if $\bm{i} = \alpha \bm{t} + \alpha' \bm{t}'$ is a linear combination of non-collinear $\bm{t}$ and $\bm{t}'$ with \emph{positive} coefficients $\alpha$ and $\alpha'$ (properly scaled such that $\lVert \bm{i} \rVert = 1$). The proof of these claims is deferred to the Appendix. In the former situation, i.e., $\bm{t} = \bm{0}$ (or $\bm{t}' = \bm{0}$), clearly the right-hand sides of \eqref{c_bis}--\eqref{c_prime_bis} cannot be equal due to $\bm{t}$ (or $\bm{t}'$) not being collinear with $\bm{i}$. In the latter situation, let us insert $\bm{i} = \alpha \bm{t} + \alpha' \bm{t}'$ into the above expressions~\eqref{c} and \eqref{c_prime} and set $\bm{c}' = \bm{c}$, then \eqref{c}--\eqref{c_prime} turn into
			\begin{subequations}
			\begin{IEEEeqnarray}{rCl}
				\bm{c}
				&=& \frac{ \bigl( \gamma_{\bm{c}}(\mathcal{I}) \alpha + \gamma_{\bm{c}}(\tilde{\mathcal{D}} \setminus \tilde{\mathcal{D}}') \bigr) \bm{t} + \gamma_{\bm{c}}(\mathcal{I}) \alpha' \bm{t}' }{\lVert \bigl( \gamma_{\bm{c}}(\mathcal{I}) \alpha + \gamma_{\bm{c}}(\tilde{\mathcal{D}} \setminus \tilde{\mathcal{D}}') \bigr) \bm{t} + \gamma_{\bm{c}}(\mathcal{I}) \alpha' \bm{t}' \rVert}   \IEEEeqnarraynumspace \label{c_bis} \\
				\bm{c}
				&=& \frac{ \gamma_{\bm{c}}(\mathcal{I}) \alpha \bm{t} + \bigl( \gamma_{\bm{c}}(\mathcal{I}) \alpha' + \gamma_{\bm{c}}(\tilde{\mathcal{D}}' \setminus \tilde{\mathcal{D}}) \bigr) \bm{t}' }{\lVert \gamma_{\bm{c}}(\mathcal{I}) \alpha \bm{t} + \bigl( \gamma_{\bm{c}}(\mathcal{I}) \alpha' + \gamma_{\bm{c}}(\tilde{\mathcal{D}}' \setminus \tilde{\mathcal{D}}) \bigr) \bm{t}' \rVert}.   \IEEEeqnarraynumspace \label{c_prime_bis}
			\end{IEEEeqnarray}
			\end{subequations}
\end{itemize}
However, these two equations cannot hold simultaneously: given that the left-hand sides of \eqref{c_bis}--\eqref{c_prime_bis} are identical, the coefficients of the linearly independent vectors $\bm{t}$ and $\bm{t}'$ on the respective right-hand sides of~\eqref{c_bis}--\eqref{c_prime_bis} would need to be identical too. Yet the ratio between the coefficients of $\bm{t}$ and $\bm{t}'$ is larger for~\eqref{c_bis} than it is for~\eqref{c_prime_bis}:
			\begin{equation*}
				\frac{ \gamma_{\bm{c}}(\mathcal{I}) \alpha + \gamma_{\bm{c}}(\tilde{\mathcal{D}} \setminus \tilde{\mathcal{D}}') }{ \gamma_{\bm{c}}(\mathcal{I}) \alpha' }
				>
				\frac{ \gamma_{\bm{c}}(\mathcal{I}) \alpha }{ \gamma_{\bm{c}}(\mathcal{I}) \alpha' + \gamma_{\bm{c}}(\tilde{\mathcal{D}}' \setminus \tilde{\mathcal{D}}) }.
			\end{equation*}
			This inequality is strict because $\tilde{\bm{v}}_n \neq \tilde{\bm{v}}_n'$ implies that at least one of the two terms $\gamma_{\bm{c}}(\tilde{\mathcal{D}} \setminus \tilde{\mathcal{D}}')$ and $\gamma_{\bm{c}}(\tilde{\mathcal{D}}' \setminus \tilde{\mathcal{D}})$ is positive.
			Hence, at least one of the two equalities~\eqref{c_bis} and \eqref{c_prime_bis} fails to hold, so we arrive at a contradiction. We can thus conclude that $\bm{c}$ and $\bm{c}'$ are \emph{not} collinear and must therefore be distinct. The self-map $\bm{\zeta}(\cdot ; \cdot)$ is thus injective. To express this in more generality, we can conclude that any $(n{-}1)$-tuple $\mathcal{V}_{\setminus i, \setminus j}$ of vertices belonging to an $n$-tuple $\mathcal{V}_{\setminus i}$, in conjunction with the locus $\bm{c}$ of its Gaussian centroid, will fully determine the $j$-th vertex $\bm{v}_j$.

\subsection{Assembling it all: tessellating $\mathbb{S}$ with spherical simplices}

We can now finally rely on two necessary conditions for a pair $(\mathcal{W}^\ast, \mathcal{V}^\ast)$ to be optimal:
\begin{itemize}
	\item	Two codewords $\bm{w}_i^\ast$ and $\bm{w}_j^\ast$ are symmetric images of each other along the $(n{-}1)$-dimensional plane passing through the origin and spanned by the vectors in $\mathcal{V}_{\setminus i, \setminus j}^\ast$.
	\item	Given $n-1$ vertices $\mathcal{V}_{\setminus i, \setminus j}^\ast$ (which are necessarily common vertices of the two tuples $\mathcal{V}_{\setminus i}^\ast$ and $\mathcal{V}_{\setminus j}^\ast$) and given that the Gaussian centroids of $\mathcal{V}_{\setminus i}^\ast$ and $\mathcal{V}_{\setminus j}^\ast$, namely $\bm{w}_i^\ast$ and $\bm{w}_j^\ast$, are also given, the remaining vertices $\bm{v}_i^\ast$ and $\bm{v}_j^\ast$ are fully determined.
\end{itemize}
Combining these two conditions, and owing to the fact that $\bm{\zeta}$ is invariant to isometry, one ends up with a third condition as a consequence: namely, that any two spherical $n$-simplices like those generated by $\mathcal{V}_{\setminus i}^\ast$ and $\mathcal{V}_{\setminus j}^\ast$ are mutually symmetric images (hence congruent).

We can now argue that $\mathcal{V}^\ast$ must be a regular simplex as follows: consider any two adjacent edges of the $(n{+}1)$-simplex corresponding to $\mathcal{V}^\ast$, say for example the segments $\bm{v}_k^\ast\bm{v}_\ell^\ast$ and $\bm{v}_\ell^\ast\bm{v}_m^\ast$ (with $k \neq m$). Consider the $(n{-}1)$-simplex $\mathcal{V}_{\setminus k, \setminus m}^\ast$ (which includes the vertex $\bm{v}_\ell^\ast$). The two $n$-simplices $\mathcal{V}_{\setminus k}^\ast$ and $\mathcal{V}_{\setminus m}^\ast$ must be symmetric along the plane spanned by the common vertices $\mathcal{V}_{\setminus k, \setminus m}^\ast$. Hence the segments $\bm{v}_k^\ast\bm{v}_\ell^\ast$ and $\bm{v}_\ell^\ast\bm{v}_m^\ast$ are mutually symmetric along that plane, and therefore of equal lengths $\lVert \bm{v}_k^\ast - \bm{v}_\ell^\ast \rVert = \lVert \bm{v}_\ell^\ast - \bm{v}_m^\ast \rVert$. Since this argument applies to any pair of adjacent edges, this means that all edges of the $(n{+}1)$-simplex $\mathcal{V}^\ast$ have equal length, hence $\mathcal{V}^\ast$ is a regular $n$-simplex. By symmetry and isometry of the centroid function, $\mathcal{W}^\ast$ must be a regular simplex too. This concludes the proof of the Weak Simplex Conjecture.

\section{Appendix}

In the following, we prove two claims from Section~\ref{ssec:injectivity}:
\begin{enumerate}
	\item	$\bm{t} = \bm{0}$ (resp. $\bm{t}' = \bm{0}$) implies non-collinearity of $\bm{i}$ and $\bm{t}'$ (resp. $\bm{i}$ and $\bm{t}$);
	\item	$\bm{i}$, $\bm{t}$, $\bm{t'}$ can only be linearly dependent if $\bm{t}$ and $\bm{t}'$ are non-collinear and $\bm{i} = \alpha \bm{t} + \alpha' \bm{t}'$ with \emph{positive} $\alpha$ and $\alpha'$.
\end{enumerate}
Recall that $\tilde{\mathcal{V}}$ and $\tilde{\mathcal{V}}'$ denote linearly independent $n$-tuples of vertices that differ only on the first element ($\tilde{\bm{v}}_1 \neq \tilde{\bm{v}}_1'$).
The cones $\tilde{\mathcal{D}}$ and $\tilde{\mathcal{D}}'$ can be expressed as intersections $\tilde{\mathcal{D}} = \bigcap_{i=1}^n \mathcal{H}_i$ and $\tilde{\mathcal{D}}' = \bigcap_{i=1}^n \mathcal{H}_i'$ of half spaces defined as
\begin{subequations}
\begin{IEEEeqnarray}{rCl}
	\mathcal{H}_i
	&=& \bigl\{ \bm{x} \in \mathbb{R}^n \colon \langle \bm{e}_i, \bm{x} \rangle > 0 \bigr\},
	\quad i = 1, \dotsc, n
	\IEEEeqnarraynumspace \\
	\mathcal{H}_i'
	&=& \bigl\{ \bm{x} \in \mathbb{R}^n \colon \langle \bm{e}_i', \bm{x} \rangle > 0 \bigr\},
	\quad i = 1, \dotsc, n
	\IEEEeqnarraynumspace
\end{IEEEeqnarray}
\end{subequations}
where $\bm{e}_i$ (resp.~$\bm{e}_i'$) are unit vectors, each of which is chosen orthogonal to the $(n-1)$-dimensional subspace spanned by $\tilde{\mathcal{V}}_{\setminus i}$ (resp.~$\tilde{\mathcal{V}}_{\setminus i}'$) and the orientation of $\bm{e}_i$ (resp.~$\bm{e}_i'$) is chosen such that $\langle \bm{e}_i, \tilde{\bm{v}}_i \rangle > 0$ (resp.~$\langle \bm{e}_i', \tilde{\bm{v}}_i \rangle > 0$). Note that $\bm{e}_1 = \bm{e}_1'$ due to common entries in $\tilde{\mathcal{V}}$ and $\tilde{\mathcal{V}}'$, so $\mathcal{H}_1 = \mathcal{H}_1'$.

Note that every triple $(\bm{e}_1, \bm{e}_i, \bm{e}_i')$ for some $i \in [2{:}n]$ is linearly dependent.
To see this, recall that from their definition, each of the three vectors $\bm{e}_1$, $\bm{e}_i$ and $\bm{e}_i'$ is orthogonal to each of the $n-2$ linearly independent vectors in $\tilde{\mathcal{V}}_{\setminus 1, \setminus i}$, whose null space is two-dimensional, hence the three vectors $\bm{e}_1$, $\bm{e}_i$ and $\bm{e}_i'$ must be coplanar (linearly dependent), satisfying $\lambda_1 \bm{e}_1 + \lambda_i \bm{e}_i + \lambda_i' \bm{e}_i' = \bm{0}$ for some coefficients $\lambda_1$, $\lambda_i$ and $\lambda_i'$ not all equal to zero. In fact, it turns out that all three coefficients are non-zero, because $\bm{e}_1$, $\bm{e}_i$, $\bm{e}_i'$ are pairwise linearly independent due to $\tilde{\mathcal{V}}$ being linearly independent (hence all null spaces of $\tilde{\mathcal{V}}_{\setminus 1, \setminus i}$ are distinct for $i \in [1{:}n]$). Let us therefore set $\lambda_1 = 1$ without loss of generality.

Upon applying the scalar product with $\tilde{\bm{v}}_i$ on both sides of the identity $\bm{e}_1 + \lambda_i \bm{e}_i + \lambda_i' \bm{e}_i' = \bm{0}$ and recalling that $\langle \bm{e}_1, \tilde{\bm{v}}_i \rangle = 0$, we get $\lambda_i \langle \bm{e}_i, \tilde{\bm{v}}_i \rangle + \lambda_i' \langle \bm{e}_i', \tilde{\bm{v}}_i \rangle = 0$. Since the scalar products $\langle \bm{e}_i, \tilde{\bm{v}}_i \rangle$ and $\langle \bm{e}_i', \tilde{\bm{v}}_i \rangle$ are both positive, it follows that $\lambda_i$ and $\lambda_i'$ must be of opposite signs (i.e., $\lambda_i \lambda_i' < 0$), so that for each $i \in [2{:}n]$ we have one of two situations: either $\lambda_1 = 1$, $\lambda_i > 0$ and $\lambda_i' < 0$, else $\lambda_1 = 1$, $\lambda_i < 0$ and $\lambda_i' > 0$. This allows us to partition the index set $[2{:}n]$ into a subset $\mathcal{J}$ and its complement $\bar{\mathcal{J}} = [2{:}n] \setminus \mathcal{J}$, where
\begin{IEEEeqnarray}{rCl}
	\mathcal{J}
	&=& \bigl\{ i \in [2{:}n] \colon \lambda_i > 0, \lambda_i' < 0 \bigr\}.
\end{IEEEeqnarray}
If $i \in \mathcal{J}$, then any $\bm{x} \in \mathbb{R}^n$ satisfying $\langle \bm{e}_1, \bm{x} \rangle > 0$ (i.e., $\bm{x} \in \mathcal{H}_1 = \mathcal{H}_1'$) and $\langle \bm{e}_i, \bm{x} \rangle > 0$ (i.e., $\bm{x} \in \mathcal{H}_i$) will also satisfy
\begin{equation*}
	\langle \bm{e}_i', \bm{x} \rangle
	= -\frac{1}{\lambda_i'} \langle \bm{e}_1, \bm{x} \rangle - \frac{\lambda_i}{\lambda_i'} \langle \bm{e}_i, \bm{x} \rangle
	> 0,
\end{equation*}
that is to say, $\bm{x} \in \mathcal{H}_i'$.
In other words, if $i \in \mathcal{J}$, then $\mathcal{H}_1 \cap \mathcal{H}_i \subset \mathcal{H}_1 \cap \mathcal{H}_i'$. Conversely, if $i \in \bar{\mathcal{J}}$, the inclusion holds in reverse, i.e., $\mathcal{H}_1 \cap \mathcal{H}_i \supset \mathcal{H}_1 \cap \mathcal{H}_i'$. Let $\mathcal{K}_i$ and $\mathcal{K}_i'$ be shorthand for $\mathcal{H}_1 \cap \mathcal{H}_i$ and $\mathcal{H}_1 \cap \mathcal{H}_i'$, respectively, so $\tilde{\mathcal{D}} = \bigcap_{i=2}^n \mathcal{K}_i$ and $\tilde{\mathcal{D}}' = \bigcap_{i=2}^n \mathcal{K}_i'$. Furthermore, $\mathcal{K}_i \subset \mathcal{K}_i'$ (for $i \in \mathcal{J}$) and $\mathcal{K}_i \supset \mathcal{K}_i'$ (for $i \in \bar{\mathcal{J}}$). It follows from these inclusion properties that
\begin{subequations}
\begin{IEEEeqnarray}{rCl}
	\tilde{\mathcal{D}} \cap \tilde{\mathcal{D}}'
	&=& \left( \textstyle\bigcap_{i \in \mathcal{J}} \mathcal{K}_i \right) \cap \left( \textstyle\bigcap_{i \in \bar{\mathcal{J}}} \mathcal{K}_i' \right)   \label{intersection}\IEEEeqnarraynumspace\\
	\tilde{\mathcal{D}} \setminus \tilde{\mathcal{D}}'
	&\subset& \left( \textstyle\bigcap_{i \in \mathcal{J}} \mathcal{K}_i \right) \setminus \left( \textstyle\bigcap_{i \in \bar{\mathcal{J}}} \mathcal{K}_i' \right)   \label{set_difference_1}\IEEEeqnarraynumspace\\
	\tilde{\mathcal{D}}' \setminus \tilde{\mathcal{D}}
	&\subset& \left( \textstyle\bigcap_{i \in \bar{\mathcal{J}}} \mathcal{K}_i' \right) \setminus \left( \textstyle\bigcap_{i \in \mathcal{J}} \mathcal{K}_i \right).   \label{set_difference_2}\IEEEeqnarraynumspace
\end{IEEEeqnarray}
\end{subequations}
To see why~\eqref{set_difference_1} holds, consider
\begin{IEEEeqnarray*}{rCl}
	\tilde{\mathcal{D}} \setminus \tilde{\mathcal{D}}'
	&=& \left( \textstyle\bigcap_{i=2}^n \mathcal{K}_i \right) \setminus \left( \left( \textstyle\bigcap_{i \in \bar{\mathcal{J}}} \mathcal{K}_i' \right) \cap \left( \textstyle\bigcap_{i \in \mathcal{J}} \mathcal{K}_i' \right) \right) \\
	&\subset& \left( \textstyle\bigcap_{i \in \mathcal{J}} \mathcal{K}_i \right) \setminus \left( \left( \textstyle\bigcap_{i \in \bar{\mathcal{J}}} \mathcal{K}_i' \right) \cap \left( \textstyle\bigcap_{i \in \mathcal{J}} \mathcal{K}_i' \right) \right) \\
	&=& \left( \textstyle\bigcap_{i \in \mathcal{J}} \mathcal{K}_i \right) \setminus \left( \textstyle\bigcap_{i \in \bar{\mathcal{J}}} \mathcal{K}_i' \right)
\end{IEEEeqnarray*}
where the last equality follows from $\textstyle\bigcap_{i \in \mathcal{J}} \mathcal{K}_i \subset \textstyle\bigcap_{i \in \mathcal{J}} \mathcal{K}_i'$. Property~\eqref{set_difference_2} is derived similarly.
The relations~\eqref{intersection}--\eqref{set_difference_2} imply that $\bm{i} \in \tilde{\mathcal{D}} \cap \tilde{\mathcal{D}}'$, $\bm{t} \in \tilde{\mathcal{D}} \setminus \tilde{\mathcal{D}}'$ and $\bm{t}' \in \tilde{\mathcal{D}}' \setminus \tilde{\mathcal{D}}$ satisfy
\begin{subequations}
\begin{IEEEeqnarray}{rClCrCl}
	\langle \bm{e}_i, \bm{i} \rangle &>& 0 \ \text{for all $i \in \mathcal{J}$}
	&,\ &
	\langle \bm{e}_i', \bm{i} \rangle &>& 0 \ \text{for all $i \in \bar{\mathcal{J}}$}   \label{scalar_product_intersection} \\
	\langle \bm{e}_i, \bm{t} \rangle &>& 0 \ \text{for all $i \in \mathcal{J}$}
	&,\ &
	\langle \bm{e}_i', \bm{t} \rangle &\leq& 0 \ \text{for some $i \in \bar{\mathcal{J}}$}   \label{scalar_product_set_difference_1} \\
	\langle \bm{e}_i', \bm{t}' \rangle &>& 0 \ \text{for all $i \in \bar{\mathcal{J}}$}
	&,\ &
	\langle \bm{e}_i, \bm{t}' \rangle &\leq& 0 \ \text{for some $i \in \mathcal{J}$.}   \IEEEeqnarraynumspace\label{scalar_product_set_difference_2}
\end{IEEEeqnarray}
\end{subequations}
provided that $\bm{t} \neq \bm{0}$ and $\bm{t}' \neq \bm{0}$ [cf.~\eqref{def:t}--\eqref{def:t_prime}]. In the contrary case, if $\bm{t} = \bm{0}$ (resp. $\bm{t}' = \bm{0}$), the conditions \eqref{scalar_product_set_difference_2} (resp. \eqref{scalar_product_set_difference_1}) and \eqref{scalar_product_intersection} render collinearity of $\bm{i}$ and $\bm{t}'$ (resp. $\bm{i}$ and $\bm{t}'$) impossible. This establishes the first claim.

If $\bm{t}$ and $\bm{t}'$ are both non-zero, we can immediately conclude that $\bm{i}$, $\bm{t}$ and $\bm{t}'$ are pairwise non-collinear, for otherwise, \eqref{scalar_product_intersection}--\eqref{scalar_product_set_difference_2} could not hold simultaneously.
Supposing that $\bm{i} = \alpha \bm{t} + \alpha' \bm{t}'$, we can outrule that $\alpha$ and $\alpha'$ are both negative, because $\bm{i}$, $\bm{t}$ and $\bm{t}'$ belong to the same half space $\mathcal{H}_1$. Also, $\alpha\alpha' = 0$ is impossible because when contrasting \eqref{scalar_product_set_difference_1}--\eqref{scalar_product_set_difference_2} against \eqref{scalar_product_intersection}, we see $\bm{i}$ cannot be collinear to either $\bm{t}$ or $\bm{t}'$. We can also outrule $\alpha\alpha' < 0$ because by~\eqref{scalar_product_set_difference_1}--\eqref{scalar_product_set_difference_2} there exist indices $j \in \bar{\mathcal{J}}$ and $k \in \mathcal{J}$ such that $\langle \bm{e}_j', \bm{t} \rangle \leq 0$, $\langle \bm{e}_j', \bm{t}' \rangle > 0$ and $\langle \bm{e}_k, \bm{t}' \rangle \leq 0$, $\langle \bm{e}_k, \bm{t} \rangle > 0$. Hence, there exists an index $i$ (equal to either $j$ or $k$) such that either $\langle \bm{e}_i, \bm{i} \rangle = \alpha \langle \bm{e}_i, \bm{t} \rangle + \alpha' \langle \bm{e}_i, \bm{t}' \rangle$ or $\langle \bm{e}_i', \bm{i} \rangle = \alpha \langle \bm{e}_i', \bm{t} \rangle + \alpha' \langle \bm{e}_i', \bm{t}' \rangle$ is negative, whereby~\eqref{scalar_product_intersection} is violated. This establishes the second claim.

\section*{Acknowledgments}

The author is indebted to Aaron Goldsmith, Alexey Balitskiy and Roman Karasev, who have pointed out some issues in an earlier draft. Helpful discussions with them to improve the manuscript, as well as with Michael Gastpar, are gratefully acknowledged.

\bibliographystyle{IEEEtran}
\bibliography{IEEEfull,references}

\end{document}

%% file: tikz/3d-sphere.tex
\tdplotsetmaincoords{0}{0}

\begin{tikzpicture}[scale=1.6,tdplot_main_coords]

\shadedraw[tdplot_screen_coords, ball color=blue, line width=0, opacity=0.2, draw=none] (0,0) circle (1);
\coordinate (O) at (0,0,0);

\coordinate (w1) at (  0.050000,  0.080000,  0.995540 ) ;
\coordinate (w2) at (  0.685273, -0.099641, -0.721438 ) ;
\coordinate (w3) at ( -0.650000, -0.575649, -0.496113 ) ;
\coordinate (w4) at ( -0.225596,  0.900000, -0.372970 ) ;

\coordinate (v1) at ( -0.2138,  0.1421, -0.9665 ) ;
\coordinate (v2) at ( -0.9155,  0.2361,  0.3258 ) ;
\coordinate (v3) at (  0.7580,  0.6154,  0.2161 ) ;
\coordinate (v4) at (  0.3609, -0.9043,  0.2281 ) ;

\coordinate (v12) at ( -0.9610,  0.2764, 0 ) ;
\coordinate (v13) at (  0.7392,  0.6735, 0 ) ;
\coordinate (v14) at (  0.3358, -0.9419, 0 ) ;

\filldraw[black!40] (v1) circle (0.2mm) node[above]{\small $v_1$};
\draw (v2) circle (0.2mm) node[left]{$v_2$};
\draw (v3) circle (0.2mm) node[above right]{$v_3$};
\draw (v4) circle (0.2mm) node[below]{$v_4$};

\filldraw (w1) circle (0.2mm) node[above]{$w_1$};
\filldraw[black!40] (w2) circle (0.2mm) node[above]{$w_2$};
\filldraw[black!40] (w3) circle (0.2mm) node[above right]{$w_3$};
\filldraw[black!40] (w4) circle (0.2mm) node[above left]{$w_4$};

\tdplotdefinepoints(0,0,0)( -0.9155, 0.2361, 0.3258 )( 0.7580, 0.6154, 0.2161 )
\tdplotdrawpolytopearc[densely dotted]{1}{anchor=west}{}
\tdplotdefinepoints(0,0,0)( 0.7580, 0.6154, 0.2161 )( 0.3609, -0.9043, 0.2281 )
\tdplotdrawpolytopearc[densely dotted]{1}{anchor=west}{}
\tdplotdefinepoints(0,0,0)( -0.9155, 0.2361, 0.3258 )( 0.3609, -0.9043, 0.2281 )
\tdplotdrawpolytopearc[densely dotted]{1}{anchor=west}{}

\tdplotdefinepoints(0,0,0)( -0.2138, 0.1421, -0.9665 )( -0.9610, 0.2764, 0.0001 )
\tdplotdrawpolytopearc[densely dotted, black!40]{1}{anchor=west}{}
\tdplotdefinepoints(0,0,0)( -0.9610, 0.2764, 0.0001 )( -0.9155, 0.2361, 0.3258 )
\tdplotdrawpolytopearc[densely dotted]{1}{anchor=west}{}
\tdplotdefinepoints(0,0,0)( -0.2138, 0.1421, -0.9665 )( 0.7392, 0.6735, 0.0001 )
\tdplotdrawpolytopearc[densely dotted, black!40]{1}{anchor=west}{}
\tdplotdefinepoints(0,0,0)( 0.7392, 0.6735, 0.0001 )( 0.7580, 0.6154, 0.2161 )
\tdplotdrawpolytopearc[densely dotted]{1}{anchor=west}{}
\tdplotdefinepoints(0,0,0)( -0.2138, 0.1421, -0.9665 )( 0.3358, -0.9419, 0.0001 )
\tdplotdrawpolytopearc[densely dotted, black!40]{1}{anchor=west}{}
\tdplotdefinepoints(0,0,0)( 0.3358, -0.9419, 0.0001 )( 0.3609, -0.9043, 0.2281 )
\tdplotdrawpolytopearc[densely dotted]{1}{anchor=west}{}

\end{tikzpicture}

%% file: tikz/2d-sphere_optimal_regions.tex
\tikzset{cross/.style={cross out, draw=black, fill=none, minimum size=2*(#1-\pgflinewidth), inner sep=0pt, outer sep=0pt}, cross/.default={2pt}}

\begin{tikzpicture}[font=\footnotesize, scale=1.2]
\useasboundingbox (-1.5,-1.3) rectangle (1.5,1.3) ;
\filldraw[draw=none, fill=blue, opacity=0.1] (0,0) circle (1) ; 
\coordinate (w1) at (+170:1) ;
\coordinate (w2) at (+45:1) ;
\coordinate (w3) at (-30:1) ;
\coordinate (O) at (0,0) ;
\filldraw (w1) circle (1pt) node[above left] {$\bm{w}_1$} ;
\filldraw (w2) circle (1pt) node[above right] {$\bm{w}_2$} ;
\filldraw (w3) circle (1pt) node[below right] {$\bm{w}_3$} ;
\coordinate (v1) at ({-30+0.5*(45+30)}:1) ;
\coordinate (v2) at ({+170+0.5*(190-30)}:1) ;
\coordinate (v3) at ({+45+0.5*(170-45)}:1) ;
\draw[densely dotted] (O) -- ($1.3*(v1)$) ;
\draw[densely dotted] (O) -- ($1.3*(v2)$) ;
\draw[densely dotted] (O) -- ($1.3*(v3)$) ;
\draw (v1) circle (1pt) node[above] {$\bm{v}_1$};
\draw (v2) circle (1pt) node[above] {$\bm{v}_2$};
\draw (v3) circle (1pt) node[above] {$\bm{v}_3$};
\node[cross] at (O) {} ;
\draw[shift={({-30+0.25*(45+30)+90}:0.0mm)}] ({-30+0.25*(45+30)}:0.95) -- ({-30+0.25*(45+30)}:1.05) ;
\draw[shift={({-30+0.75*(45+30)+90}:0.0mm)}] ({-30+0.75*(45+30)}:0.95) -- ({-30+0.75*(45+30)}:1.05) ;
\draw[shift={({+170+0.25*(190-30)+90}:0.2mm)}]  ({+170+0.25*(190-30)}:0.95) -- ({+170+0.25*(190-30)}:1.05) ;
\draw[shift={({+170+0.25*(190-30)+90}:-0.2mm)}]  ({+170+0.25*(190-30)}:0.95) -- ({+170+0.25*(190-30)}:1.05) ;
\draw[shift={({+170+0.75*(190-30)+90}:0.2mm)}]  ({+170+0.75*(190-30)}:0.95) -- ({+170+0.75*(190-30)}:1.05) ;
\draw[shift={({+170+0.75*(190-30)+90}:-0.2mm)}]  ({+170+0.75*(190-30)}:0.95) -- ({+170+0.75*(190-30)}:1.05) ;
\draw[shift={({+45+0.25*(170-45)+90}:-0.4mm)}]  ({+45+0.25*(170-45)}:0.95) -- ({+45+0.25*(170-45)}:1.05) ;
\draw[shift={({+45+0.25*(170-45)+90}:+0.0mm)}]  ({+45+0.25*(170-45)}:0.95) -- ({+45+0.25*(170-45)}:1.05) ;
\draw[shift={({+45+0.25*(170-45)+90}:+0.4mm)}]  ({+45+0.25*(170-45)}:0.95) -- ({+45+0.25*(170-45)}:1.05) ;
\draw[shift={({+45+0.75*(170-45)+90}:-0.4mm)}]  ({+45+0.75*(170-45)}:0.95) -- ({+45+0.75*(170-45)}:1.05) ;
\draw[shift={({+45+0.75*(170-45)+90}:+0.0mm)}]  ({+45+0.75*(170-45)}:0.95) -- ({+45+0.75*(170-45)}:1.05) ;
\draw[shift={({+45+0.75*(170-45)+90}:+0.4mm)}]  ({+45+0.75*(170-45)}:0.95) -- ({+45+0.75*(170-45)}:1.05) ;
\node at ({0.5*(-30+0.5*(45+30) + 45+0.5*(170-45))}:0.4) {$\mathcal{D}_2$} ;
\node at ({0.5*(+45+0.5*(170-45) + 170+0.5*(190-30))}:0.4) {$\mathcal{D}_1$} ;
\node at ({0.5*(+170+0.5*(190-30) -360 -30+0.5*(45+30))}:0.4) {$\mathcal{D}_3$} ;
\end{tikzpicture}

%% file: tikz/2d-sphere_relaxation.tex
\tikzset{cross/.style={cross out, draw=black, fill=none, minimum size=2*(#1-\pgflinewidth), inner sep=0pt, outer sep=0pt}, cross/.default={2pt}}

\begin{tikzpicture}[font=\footnotesize, scale=1.2]
\useasboundingbox (-1.5,-1.3) rectangle (1.5,1.3) ;
\filldraw[draw=none, fill=blue, opacity=0.1] (0,0) circle (1) ; 
\coordinate (w1) at (190:1) ;
\coordinate (w2) at (40:1) ;
\coordinate (w3) at (-30:1) ;
\coordinate (O) at (0,0) ;
\filldraw (w1) circle (1pt) node[above left] {$\bm{w}_1$} ;
\filldraw (w2) circle (1pt) node[above right] {$\bm{w}_2$} ;
\filldraw (w3) circle (1pt) node[below right] {$\bm{w}_3$} ;
\coordinate (v1) at (10:1) ;
\coordinate (v2) at (220:1) ;
\coordinate (v3) at (85:1) ;
\draw[densely dotted] (O) -- ($1.3*(v1)$) ;
\draw[densely dotted] (O) -- ($1.3*(v2)$) ;
\draw[densely dotted] (O) -- ($1.3*(v3)$) ;
\draw (v1) circle (1pt) node[above] {$\bm{v}_1$};
\draw (v2) circle (1pt) node[above] {$\bm{v}_2$};
\draw (v3) circle (1pt) node[above] {$\bm{v}_3$};
\node[cross] at (O) {} ;
\node at ({0.5*(85+220)}:0.4) {$\mathcal{D}_1$} ;
\node at ({0.5*(10+85)}:0.4) {$\mathcal{D}_2$} ;
\node at ({0.5*(220+10-360)}:0.4) {$\mathcal{D}_3$} ;
\end{tikzpicture}

%% file: tikz/cone.tex
\begin{tikzpicture}[scale=0.6]
	\filldraw[draw=none, fill=blue, opacity=0.1] (190:2cm) circle (2) ; 
	\draw (0,0) --+ (10:3cm) coordinate (v1tipp) ;
	\draw (0,0) --++ (180+10:2) coordinate (origin) --++ (160:5) node[pos=0.7, above] {$\mathcal{D}_i$} ;
	\draw[fill=black] (origin) circle (1.5pt) ;
	\draw[->] (origin) --+ (50:2cm) coordinate (ctip) node[above right] {$\bm{c}$} ;
	\draw[->] (origin) --+ (10:2cm) coordinate (v1tip) node[below] {$\bm{v}_1$} ;
	\draw[->] (origin) --+ (160:2cm) coordinate (v2tip) node[above] {$\bm{v}_2$} ;
	\draw[<-, densely dotted] (origin) -- node[above, sloped, pos=0.55] {\footnotesize $2\sigma\bm{c} u_1$} +(50+180:1.5cm) ;
	\draw (ctip) --+ (50+90:2cm) coordinate (xsegmentl) --+ (50-90:4cm) coordinate (xsegmentr);
	\begin{scope}[shift=(50+180:1.5cm)]
		\draw (10:3) coordinate (v1tippdown) -- (180+10:2) coordinate (origin) --++ (160:5) node[pos=0.7, above] {$\bar{\mathcal{D}_i}$} ;
	\end{scope}
	\coordinate (i1) at (intersection of xsegmentl--xsegmentr and v1tip--v1tipp) ;
	\coordinate (i2) at (intersection of xsegmentl--xsegmentr and v1tippdown--origin) ;
	\fill[gray, opacity=0.2] (i2) -- (i1) -- (v1tipp) -- (v1tippdown) ;
\end{tikzpicture}